\magnification=\magstep1
\parindent=15pt

\centerline{\bf New Developments on Gravitational Force}
\centerline{\bf and Nonlinear Oscillations of Space}
\vskip 14pt
\centerline{\bf Behram N. Kursunoglu}
\centerline{Global Foundation, Inc.}
\centerline{P.O. Box 249055}
\centerline{Coral Gables, Florida  33124-9055}
\centerline{e-mail:kursun@globalfoundationinc.org}
\vskip 16pt
\centerline{\bf Abstract}
\vskip 12pt
This paper contains results obtained as solutions of the Unified Field
Theory equations.  It yields space nonlinear oscillations, a quartet of
gravitational forces, quintessence, and replaces Einstein's Cosmological
Constant by an invariant parameter $r_0$ which prevails over the entire
evolution of the Universe.
\vskip 12pt
In the 1999 Coral Gables Conference I had reported on the latest
observations, with regard to the expansion of the universe with
increasing acceleration which theoretically was explained in terms of the
existence of repulsive gravitational forces predicted in 1991 from the
unified theory$^1$ (see page 493 in this reference) and further that space
itself was subjected to some nonlinear oscillations with variable frequencies.
These predictions of the unified theory were published in the
proceedings of the 1999 Coral Gables Conference$^2$.  Contrary to the
existing paradigm amongst most cosmologists and physicists, I strongly
agree with Einstein's regret for his introduction of the cosmological
constant into (though for different reasons) his general relativistic
theory of gravitation.  In the unified theory the 16 component
non-symmetric tensor ${\hat g}_{\mu v} =  g_{\mu v} + \varepsilon q^{-1}\Phi_{\mu v}$
where $\varepsilon = \pm 1, \pm i$ contains the invariant parameter $q$,
where $q^2$ is an energy density and is related to an intrinsic
fundamental length parameter $r_0$ by the relation
$$
r^2_0q^2 = c^4/2G. \eqno(1)
$$
This relation was obtained$^3$ in 1950 while a graduate student in
Cambridge University and it is the condition for the non-symmetric
theory to reduce to general relativity in the limit of $r_o=0$, the
correspondence limit$^4$.  The parameter $q$ has the dimensions of an
electric field and therefore the field tensor ${\hat g}_{\mu v}$ is
dimensionless.  The relation (1) can also be regarded as an equation of
state.  The role of $r_0$, which varies between zero and infinity, in
this theory is as important as the role of $\hbar$ in quantum theory.
This is the missing link in Einstein and Schr\"odinger versions$^5$ of
the non-symmetric theory.  Both Einstein$^6$ (see the Appendix II page
146) and Schr\"odinger$^1$ (see the letter to me in the Appendix of this
reference) have commented on the subject of constants in a physical
theory.
\vskip 12pt
The expansion of the universe is now linked up with the predicted
existence of the {\it quartet of gravitational forces}.  This
multiplicity of the gravitational forces can be seen by studying the sign
of the gravitational constant G in the equation of state (1).  The four
possible substitutions in the relation (1) for the two invariant
parameters $r_0$ and $q$ are
$$
G_1^+ = [\pm r_0,~ \pm q],~G_1^- = [\pm ir_0,~ \pm q],~G_2^- = [\pm r_0,
~\pm iq],~G_2^+ = [\pm ir_0,~ \pm iq ], \eqno (2)
$$
$G^+ = G$ (attractive gravity), and $G^- = -G$ (repulsive gravity).
Thus, the quartet of gravitational forces consists of two attractive
$G_1^+, G_2^+$ and two repulsive $G_1^-, G_2^-$ forces.
\vskip 12pt
\noindent
I would like, for the use of the field equations of unified theory, to
introduce an invariant function $f$, which was already contained in the
original field equation, and a new variable $\beta$ by the
relation
$$
dr = f d\beta. \eqno (3)
$$
\noindent
In terms of the new variable $\beta$, the spherically symmetric field
equations can be written as
$$
{1\over 2} r_0^2 {d\over d\beta} (S \exp(\rho_{n \tau}) {\dot
\Phi}_{ns \tau}) = R^2 \cos \Phi_{ns \tau} + l_0^2 \sin \Phi_{ns
\tau},\eqno (4)
$$
$$
{1\over 2} r_0^2 {d\over d\beta} (S
\exp(\rho_{n\tau}){\dot\rho}_{ns\tau})
= -R^2 \sin\Phi_{ns\tau} + l_0^2 \cos\Phi_{ns\tau} +
\exp(\rho_{n\tau}), \eqno (5)
$$
$$
{1\over 2} r_0^2 {d\over d\beta} ({\dot S} \exp(\rho_{n\tau})) =
\exp(\rho_{n\tau}) \left(1 - {\exp(\rho_{n\tau}) \sin\Phi_{ns\tau}\over
R^2 + r_0^2}\right), \eqno (6)
$$
$$
2{\ddot\rho}_{n\tau} + {\dot \rho}^2_{ns\tau} + {\dot \Phi}^2_{ns\tau}
= 0, \eqno (7)
$$
\noindent
where
$$
{\dot \rho} = {d\rho\over d\beta}, \eqno (8)
$$
\noindent
and where now $S,\rho$, and $\Phi$ can be regarded as functions of the
new variable $\beta$.  The equation (7) is derivable from the three
equations (4), (5), (6) which can be written as
$$
({d^2\over d\beta^2} + \omega^2) \exp ({{1\over 2}\rho}) = 0,~\omega =
{1\over 2}
{\dot \Phi}, \eqno (9)
$$
\noindent
implies nonlinear oscillations of the {\it orbiton} magnetic charge
layers.  Hence we see that the structure of  an orbiton, in view of the
variable frequency $\omega$, oscillates like a pendulum whose length is
changing.  An orbiton, which confines all the monopoles, is formed as a result of a
condensation$^1$ of infinite number of magnetic charges or monopoles
into alternating and decreasing thickness of the
layers of positive and negative charges (with zero sum).  At
distances large compared to $r_0$ the function $\exp({{1\over 2}
\rho})\rightarrow r$.  The functions S and R are defined by
$$
f = \upsilon \cosh\Gamma,~~S = {\exp(u)\over \cosh^2\Gamma}, \eqno (10)
$$
$$
\cosh\Gamma = (R^2 + r^2_0) \exp(-\rho), ~[(\exp (2\rho) +
\lambda_0^4)]^{1/2} = R^2 + r_0^2, \eqno (11)
$$
$$
\ell_0^2 = q^{-1} \mid g \mid, \lambda_0^2 = q^{-1} \mid e \mid, \eqno
(12)
$$
\noindent
where e and g represent electric and magnetic charges, respectively.  If
we assume $\Phi$ is a constant then there are no oscillations and we
can calculate $r_0$ in the form
$$
r^2_0 = (\ell_0^4 + \lambda_0^4)^{1/2}, ~{\rm where}~\ell_0^2 = (2G/c^4)
N^2 \mid g \mid (e^2 + g^2)^{1/2}, \eqno (13)
$$
$$
\lambda_0^2 = (2G/c^4) N^2 \mid e\mid (e^2 + g^2)^{1/2}, \eqno (14)
$$
\noindent
and $\lambda_0$ and $\ell_0$ are constants of integration.
\vskip 12pt
The dynamics of space in the presence of matter is not simple.  The
existence of magnetic charge carrying quintessence with repulsive
gravitational force versus ordinary matter with attractive gravitational
force are in a state of competition which drives the expansion of the
universe with increasing acceleration.  The presence of the space
oscillations in the vicinity of matter do provide regularity to the
expansion of the universe.  The oscillations are not confined to a point
but, as in the case of superstrings theory, they behave like a patch
where the oscillating system has, as in the general relativity, a
Schwarzschild radius.  The size of a patch can be of the order of $r_0$
and its mass can be defined as
$$
M = (c^2/2G) r_0, \eqno (15)
$$
\noindent
where $r_0$ is equal to the Schwarzschild radius of the mass M.  If we
treat a proton as a patch then its Schwarzschild radius is of the order
of $10^{-54}$cm.  The largest oscillating patch is the universe itself
with its Schwarzschild radius being equal to its size $r_0 = 10^{28}$cm.
In this case i.e., $r_0 \sim \infty$ and the field equations (4)-(7)
would yield flat space-time solutions which would indicate the fact that
universe is flat.
\vskip 12pt
In this paper I have thus shown that the fundamental parameter $r_0$
differs from Einstein's concept of cosmological constant in a
fundamental way since in this case a single parameter is part of the
non-symmetric structure and prevails over the entire evolution of the
universe.  Why did Einstein not consider the necessity and highly
visible existence of $r_0$? Was he mostly influenced by the
disappointment with his own cosmological constant?  Actually Einstein
answers this question, presumably, based on my correspondence with him
during 1950-1952 on the subject matter of constants in the field
equations, in his famous book$^6$, page 146, ``all such additional terms
bring a heterogeneity into the system of equations, and can be
disregarded, provided that no strong physical argument is found to
support them''.  I must point out that without $r_0$ we can not even
obtain the classical laws of motion of a charged particle in an
electromagnetic field.  I obtained many additional results to refute
entirely Einstein's claim quoted above.  The most striking fact can be
found in the definition of mass:  elementary particles, the sun,
the
earth, a black hole, neutron star etc.  obtained as
$$
M = (c^2/2G)r_0 = (1/2) (r_0/r_p) m_p, \eqno (16)
$$
\noindent
where $r_p$ and $m_p$ represent Planck length and Planck mass,
respectively.
\vskip 12pt
Finally, with regard to nonlinear oscillations, there is the possibility
that space itself consists of fluctuating bubbles as often referred to
as {\it quantum foam} or in this case we may prefer a description like
{\it nonlinear foam}.

\vfill\eject
\leftline{\bf References:}
\vskip 12pt
\item{(1)} Behram N. Kursunoglu, {\it After Einstein and Schr\"odinger, A New
Unified Field Theory} Journal of Physics Essays, {\bf 4}, pp439-518,
(1991).
\vskip 6pt
\item{(2)} Behram N. Kursunoglu, Coral Gables 1999 Conference Proceedings
``Confluence of Cosmology, Massive Neutrinos, Elementary Particles, and
Gravitation'' pages 5-23, Edited by Behram N. Kursunoglu, Stephan L.
Mintz, Arnold Perlmutter.
\vskip 6pt
\item{(3)} Behram N. Kursunoglu, Phys.Rev.{\bf 88}, 1369 (1952); Phys.
Rev. {\bf D9},
2723(1974).
\vskip 6pt
\item{(4)} Behram N. Kursunoglu, Phys. Rev. {\bf D13}, 1538 (1976).
\vskip 6pt
\item{(5)} Abraham Pais, {\it Subtle is the Lord.....}Science and the Life of
Albert Einstein, Oxford University Press, (1982).
\vskip 6pt
\item{(6)} Albert Einstein, {\it The Meaning of Relativity}, Princeton
University Press, (1953).
\end